\documentclass[sigconf,10pt]{acmart}
\renewcommand\footnotetextcopyrightpermission[1]{} 
\usepackage{lmodern}
\usepackage{graphicx}

\begin{document}
\title{A Survey of Distributed Message Broker Queues}

\author{Vineet John}
  \affiliation{%
    \institution{University of Waterloo}}
    \email{v2john@uwaterloo.ca}  
\author{Xia Liu}
  \affiliation{
    \institution{University of Waterloo}
  }
  \email{x397liu@uwaterloo.ca} 

\date{}

\begin{abstract}
  This paper surveys the message brokers that are in vogue today for distributed communication. Their primary goal is to facilitate the construction of decentralized topologies without single points of failure, enabling fault tolerance and high availability. These characteristics make them optimal for usage within distributed architectures. However, there are multiple protocols built to achieve this, and it would be beneficial to have a empirical comparison between their features and performance to determine their real-world applicability. 
  
  This paper focuses on two popular protocols (Kafka and AMQP) and explores the divergence in their features as well as their performance under varied testing workloads.
\end{abstract}

\maketitle

\section{Introduction}
\label{sec:introduction}
  Distributed Message Brokers are typically used to decouple separate stages of a software architecture. They permit communication between these stages asynchronously, by using the publish-subscribe paradigm.\cite{banavar1999case}. These message brokers are also finding new applications in the domain of IoT devices and may also be used as method to implement an event-driven processing architecture. 
  
  A few of the current offerings of this architectural paradigm include Apache Kafka\cite{kreps2011kafka}, AMQP\cite{vinoski2006advanced}, ActiveMQ\cite{snyderintroduction}. 

  The rest of this paper is structured as follows:
  \begin{itemize}
    \item 
    Section \ref{sec:research_questions}: Research questions investigated
    \item 
    Section \ref{sec:kafka}: Kafka - architecture and overview
    \item 
    Section \ref{sec:amqp}: AMQP - architecture and overview
    \item 
    Section \ref{sec:experimental_results}: Benchmarking results
    \item 
    Section \ref{sec:comparison}: Kafka vs. AMQP - performance comparison
    \item 
    Section \ref{sec:conclusion}: Conclusion and final remarks
    \item 
    Section \ref{sec:appendix}: Appendix
  \end{itemize}


\section{Research Questions}
\label{sec:research_questions}
  The following research questions are enumerated for further study:
  \begin{itemize}
    \item [\textbf{RQ0}] What are the message broker implementations commonly in use today?
    \item [\textbf{RQ1}] What are the common requirements for the implementation of message queues?
    \item [\textbf{RQ2}] What are the divergent functionalities in the current message queue offerings?
    \item [\textbf{RQ3}] How do each of the implementations offer reliability, partitioning and fault tolerance?
  \end{itemize}

\section{Kafka}
\label{sec:kafka}

  Kafka was developed at LinkedIn and primarily used for log processing. This worked well for Kafka's user engagement metrics collection use-case. The fundamental features behind Kafka are performance over reliability and it offers high throughput, low latency message queuing. Kafka can be used either for online or offline log processing.

  The fact that reliability of message reception is traded off to some extent implies the loss of a single record among a multitude is not a huge deal-breaker. The rationale behind this is, for log aggregated data, delivery guarantees are unnecessary. 

  \begin{figure} 
  \centering
    \includegraphics[width=.5\textwidth]{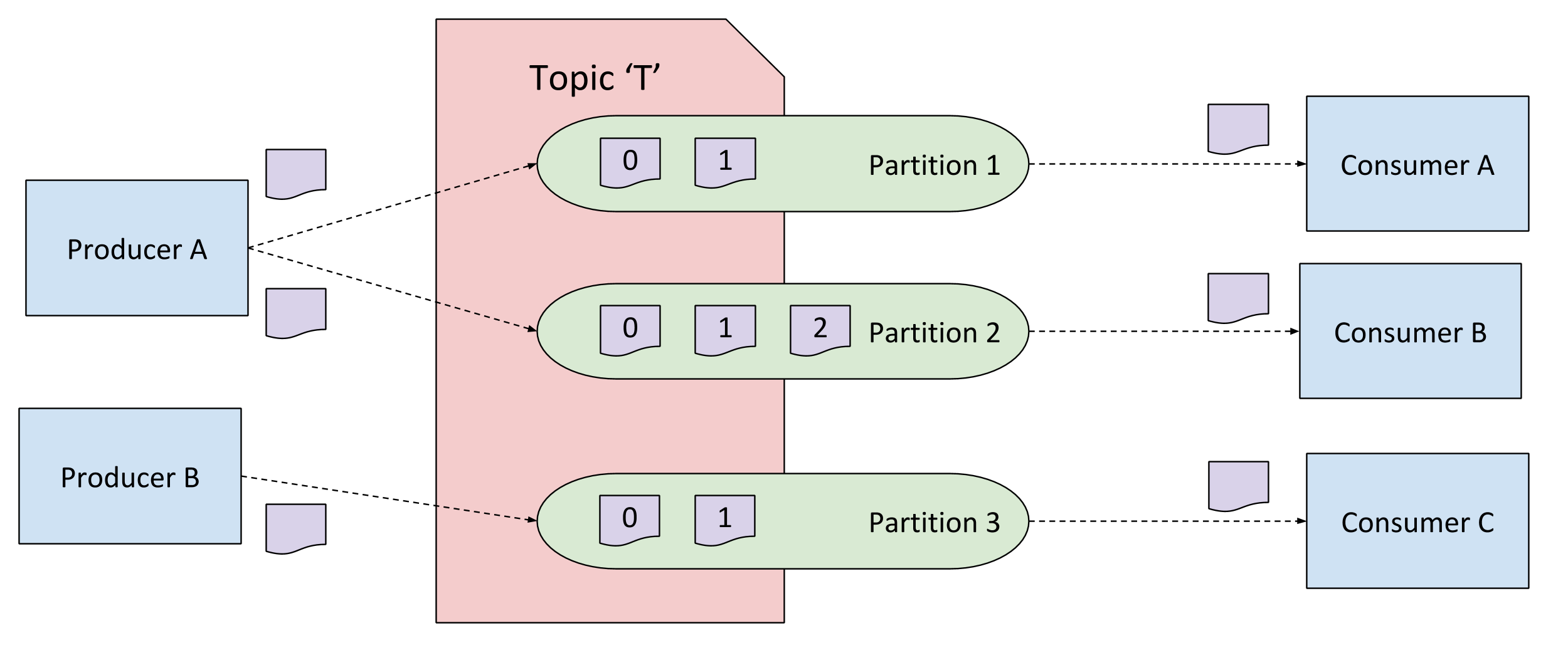}
    \caption{Kafka Architecture}
  \label{fig:kafka-architecture}  
  \end{figure}

  \subsection{Kafka - Architecture} 
  \label{sub:kafka_architecture}
    The general architecture for Kafka is illustrated in Figure \ref{fig:kafka-architecture}.

    Data is divided into topics, which resemble streams of messages. Topics are further divided into partitions and each broker can possess one or more such partitions. Producers publish to topics and brokers store messages received from the producers. Messages are payloads of bytes that consumers use iterators to consume. Kafka employs the pull model for consumer message receipt, which permits consumers to consume messages at their own pace. 

    Kafka supports 2 models for topic subscription:
    \begin{itemize}
      \item
      Point-to-point: A single message copy is read by an arbitrary consumer.
      \item
      Pub-sub: Each consumer has it's own copy of each message.
    \end{itemize}
  


\section{AMQP}
\label{sec:amqp}
  It is an asynchronous message queuing protocol, aiming to create an open standard for passing messages between applications and systems regardless of internal design. It was initially designed for financial transaction processing systems, such as trading and banking systems, which require high guarantees of reliability, scalability, and manageability. This use-case greatly influences the design of AMQP.

  \begin{figure}[ht]
      \centering
      \includegraphics[width=0.5\textwidth]{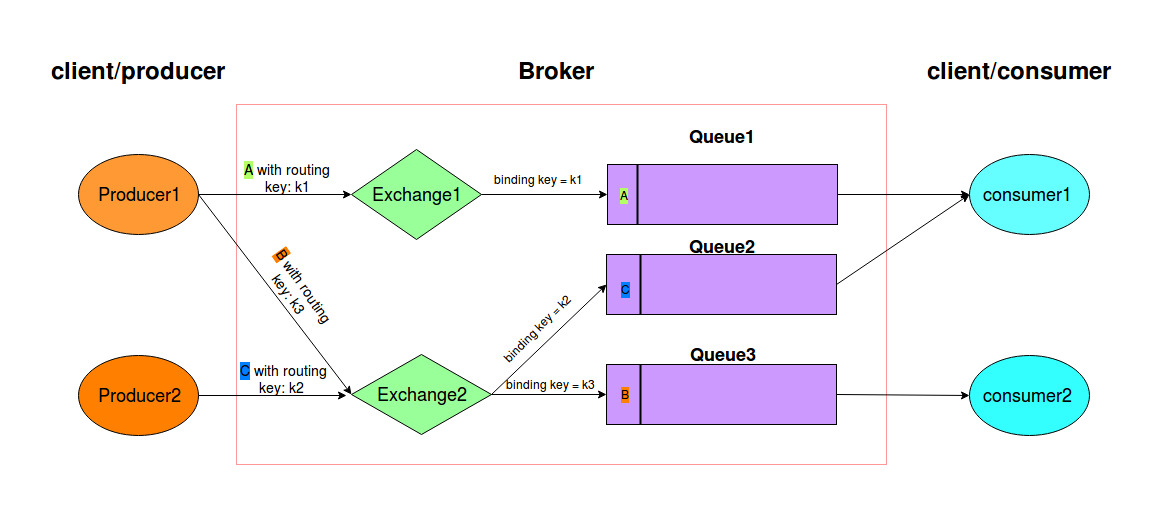}
      \caption{AMQP Workflow}
      \label{fig:amqp-architecture} 
  \end{figure}

  \subsection{AMQP Architecture}
  \label{sub:amqp-architecture}
    Figure \ref{fig:amqp-architecture} shows the architecture of AMQP.  There are several concepts in AMQP need to be listed and explained before illustrating how an AMQP broker works. These terms are elaborated upon in the Appendix (Section \ref{sec:appendix}\ref{sub:amqp_components}).

    Instead of sending messages to queues directly, producers send messages to exchanges. Exchanges then send messages to queues whose binding key matches the messages' routing key. The queue, in turn, sends messages to consumers who are subscribed to it. This design of exchanges and queues in AMQP makes the message passing transparent. A producer does not need to maintain any state about its messages' consumers. Similarly, a consumer does not need to maintain state about the producers. The decisions of which queue the message should be routed to is decided by the exchange instead of the application, which makes changing message-routing and delivery logic easier compared to a messaging system where the application makes such decisions.


\section{Experimental Results}
\label{sec:experimental_results}

  \subsection{Testbed}
  \label{sub:testbed}
    The testbed utilized to perform the benchmarking is comprised of 5 nodes, each with a similar hardware configuration: 12 cores @ 2300.13Mhz, 16GB memory, with an HDD for persistent storage, and a network interface capacity of 1Gbps.

    There were 2 types of tests run for Kafka and RabbitMQ (which is the implementation used to proxy the behavior of AMQP). The single producer/consumer test keeps the overall test workload constant (1 million messages, 50B each) and scales the queue deployment from 1 to 5 nodes. The multiple producer/consumer setup keeps the number of nodes constant and scales the number of producers/consumers connecting from each node. 
    
    All the benchmarks were run using a modified version of Flotilla\footnote{https://github.com/v1n337/flotilla}, which is a message broker benchmarking tool written in Go.

    Figures \ref{fig:kakfa-throughput-singlepc}, \ref{fig:kakfa-throughput-multiplepc}, \ref{fig:kakfa-latency-singlepc} and \ref{fig:kakfa-latency-multiplepc} illustrate the throughput and latency of Kafka under the aforementioned workloads. Similar benchmark results for RabbitMQ are shown in figures \ref{fig:rabbitmq-throughput-singlepc}, \ref{fig:rabbitmq-throughput-multiplepc} and \ref{fig:rabbitmq-latency}.

  \subsection{Benchmarking Results}
    \begin{figure} 
    \centering
      \includegraphics[width=.4\textwidth]{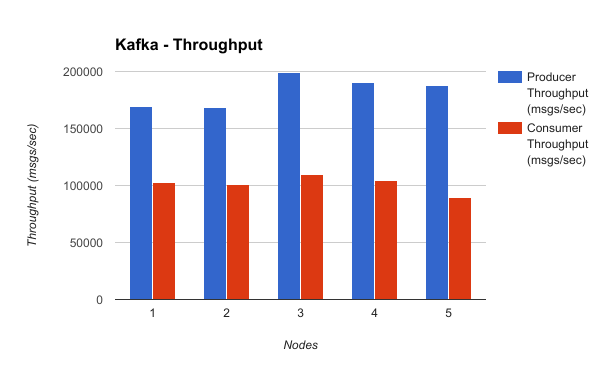}
      \caption{Kafka Throughput (Single Producer/Consumer per node)}
    \label{fig:kakfa-throughput-singlepc}  
    \end{figure}

    \begin{figure} 
    \centering
      \includegraphics[width=.4\textwidth]{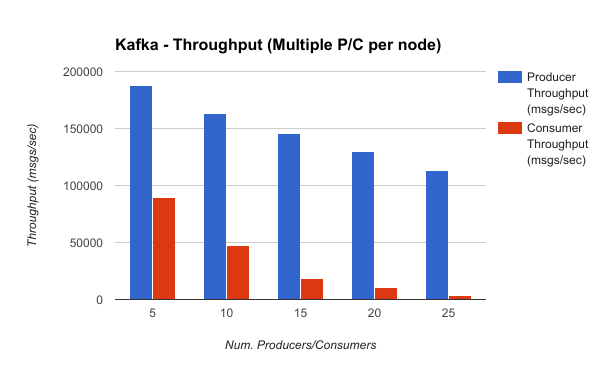}
      \caption{Kafka Throughput (Multiple Producers/Consumers per node)}
    \label{fig:kakfa-throughput-multiplepc}  
    \end{figure}

    \begin{figure} 
    \centering
      \includegraphics[width=.4\textwidth]{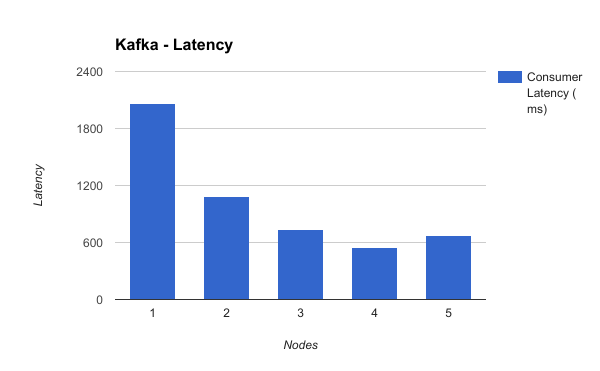}
      \caption{Kafka Latency (Single Producer/Consumer per node)}
    \label{fig:kakfa-latency-singlepc}  
    \end{figure}

    \begin{figure} 
    \centering
      \includegraphics[width=.4\textwidth]{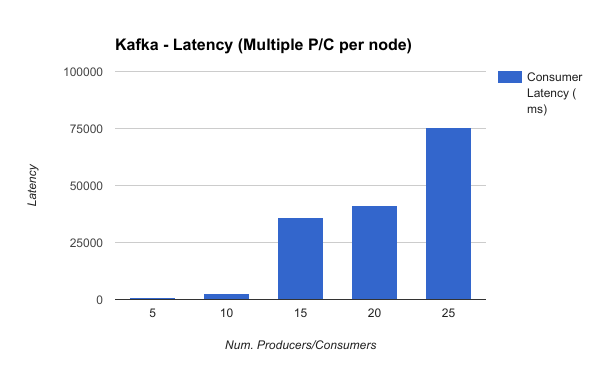}
      \caption{Kafka Latency (Multiple Producers/Consumers per node)}
    \label{fig:kakfa-latency-multiplepc}  
    \end{figure}

    \begin{figure} 
    \centering
      \includegraphics[width=.4\textwidth]{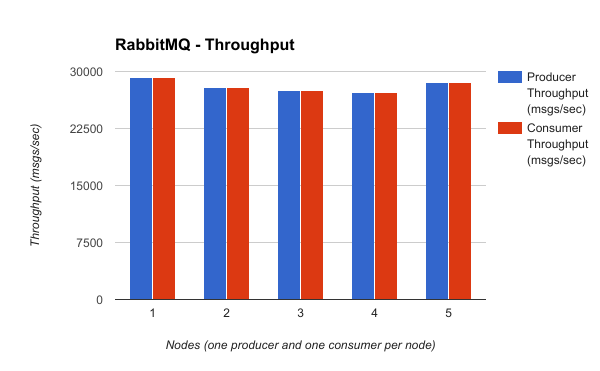}
      \caption{RabbitMQ Throughput (Single Producer/Consumer per node)}
    \label{fig:rabbitmq-throughput-singlepc}  
    \end{figure}

    \begin{figure} 
    \centering
      \includegraphics[width=.4\textwidth]{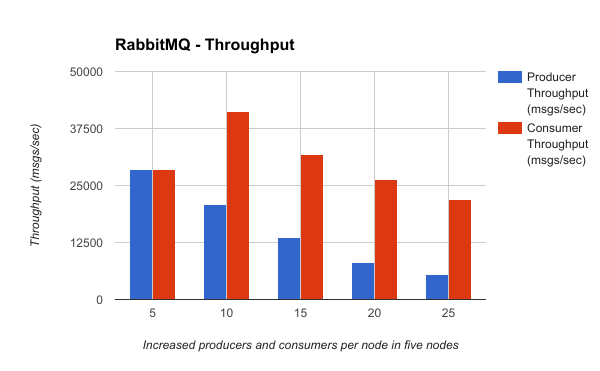}
      \caption{RabbitMQ Throughput (Multiple Producers/Consumers per node)}
    \label{fig:rabbitmq-throughput-multiplepc}  
    \end{figure}

    \begin{figure} 
    \centering
      \includegraphics[width=.4\textwidth]{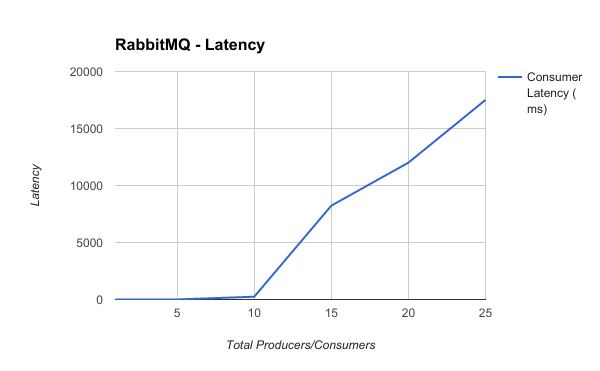}
      \caption{RabbitMQ Latency}
    \label{fig:rabbitmq-latency}  
    \end{figure}

    The results of benchmarking Kafka while keeping the same payload and scaling to multiple nodes (Fig. \ref{fig:kakfa-throughput-singlepc}, Fig. \ref{fig:kakfa-latency-singlepc}) shows a latency drop by a factor of about 3x, in comparison to only a nominal penalty on throughput (1.06x). This just goes to demonstrate the effectiveness of Kafka being deployed in a cluster setup in handling large log-aggregation-like workloads.

    The corresponding results while keeping the number of nodes constant and increasing the payload by using additional producers and consumers causes the performance metrics to degrade in a more pronounced manner (Fig. \ref{fig:kakfa-throughput-multiplepc}, Fig. \ref{fig:kakfa-latency-multiplepc}). In particular, the latency increases by about 2 orders of magnitude with a 5x increase in payload coming from a 5x increase in publishers and consumers. This also corroborates the 29x drop in consumer throughput, meaning that messages stay in the Kafka log partitions (queues) a lot longer and remain unconsumed, than if there was only a single publisher/subscriber per node. This is especially visible due to the huge disparity between the producer throughput and the consumer throughput. This could be attributed to an increased number of accesses of the metadata layer i.e. Zookeeper becoming a focus point of resource contention on a single node, causing context switching and resulting in additional overhead.

    Figure \ref{fig:rabbitmq-throughput-singlepc} describes RabbitMQ clients' throughput when each node runs single producer and consumer, similar to the setup for Kafka. The reason for this stable producer throughput could be attributed to a high exchange routing speed. The rate at which the exchange processes messages is faster than the rate of producers producing and sending messages. Another cause could be that the broker is capable of matching this rate resulting in a consumer throughput that is almost the same as producer throughput. Consumers consume messages faster than producers are able to publish messages to the exchanges. Therefore, queues can be empty, and consumers have to wait for messages from producers. As a result, producer throughput is a ceiling for consumer throughput in this experiment.
    
    Figure \ref{fig:rabbitmq-throughput-multiplepc} shows RabbitMQ clients' throughput with the rapid growth of clients. In general, scaling the number of producers and consumers resulting in a reduction in throughput, while consumers throughput is not limited by a single producer anymore and can consume messages sourced from multiple producers at a higher rate. The overall reduction in throughput can be attributed to resource contention, similar to what Kafka encounters.
    
    Figure \ref{fig:rabbitmq-latency} illustrates the consumer latency with the increasing number of total clients. The latency remain low for benchmarks using a single producer and consumer per node. The main reason for low latency at the beginning and sharp increased latency is due to the queues being backed-up with messages from multiple producers as opposed to almost empty. RabbitMQ, and all distributed queues in general, operate at optimal latency when queues are relatively empty. When the number of producers increase to 15, the exchange is saturated with message traffic, and messages start queuing for longer durations, thereby increasing the latency.

\section{Comparison} 
\label{sec:comparison}

  The results of Kafka and AMQP benchmarking test reveal that Kafka has higher throughput while AMQP has lower latency. 

  There are a few potential reasons for why Kafka provides a higher throughput, explained by its inherent optimizations:
  \begin{itemize}
    \item 
    The SendFile API in Kafka helps bypass kernel buffers to write data directly from the file channel to the socket.
    \item 
    Writing to a message broker in Kafka uses sequential disk writes and page-level caching also help to maximize throughout.
    \item 
    Batching in Kafka is available out-of-the-box, and this helps minimize communication overhead.
  \end{itemize}

  In terms of latency, the consumer push-model option results better mean latency in AMQP; while Kafka uses the pull-model where consumers have to fetch messages from broker. Another reason is that AMQP's default configuration doesn't persist messages to disk at the broker, but Kafka does, resulting in a longer intermediate delay before message consumption.
    

\section{Conclusion} 
\label{sec:conclusion}
  Based on the experimental results and comparisons in sections \ref{sec:experimental_results} and \ref{sec:comparison}, a distinct contrast between each of the message queue offerings can be made.

  Two key aspects should be considered making a choice of message broker protocol, namely \textbf{throughput} and \textbf{reliability}. If reliability for the application is not critical, Kafka is the better choice, as evidenced by the experimental results. For example, for an application that needs to process a large amount of messages is tolerant to losing a few. Usage examples include page view counters, advertisement clicks and social media reactions.
  
  On the other hand, if messages are important, such as financial transactions, the cost of losing any of messages is far higher that not achieving an optimal throughput and the application is encouraged to use AMQP. In addition, AMQP can encrypt messages out-of-the-box. Therefore, applications requiring security should also consider using AMQP. Moreover, four types of exchanges in AMQP allow AMQP broker work in peer-to-peer, publisher/consumer and broadcasting models. The last two models are suitable for instant messaging and notification services.

  This concludes the study of two popular distributed message broker queuing protocols, Kafka and AMQP. Future work would entail exploring other implementations such as ZeroMQ, ActiveMQ and Qpid in depth. For the purposes of this paper, these implementations have been summarized at a high level in the Appendix (Section \ref{sec:appendix}\ref{sub:other_msg_broker_impl}).


\section{Acknowledgements} 
\label{sec:acknowledgements}
  The authors would like to thank Dr. Samer Al-Kiswany, for his guidance and supervision on this project, as well as for his constant encouragement.


\bibliographystyle{ACM-Reference-Format}
\bibliography{sigproc} 

\section{Appendix}
\label{sec:appendix}

\appendix

\section{Feature Comparison}
\label{sec:feature_comparison}
  Table \ref{tab:message-broker-comparison} summarizes, at a high-level, the differences between Kafka and AMQP. These differences illustrate that the goal for Kafka is to focus on high throughput while AMQP tries to achieve better reliability. For example, AMQP can use consumer ACKs to ensure reliability; while, Kafka's highest guarantee is an ACK for whether the message was written persistently to the broker log. This option increases AMQP reliability, but decreases its throughput. 

  \begin{table}
    \begin{tabular}{| p{0.3\linewidth} | p{0.3\linewidth} | p{0.3\linewidth} |}
      \hline 
        \textbf{Characteristic} & \textbf{Kafka} & \textbf{AMQP}\\
      \hline
        Message Format & Bytes - Easier to develop for & Binary - better compression, boosts throughput \\
      \hline
        Message Routing & No intermediaries. Messages are sent to brokers & Exchanges used to route messages using bindings \\
      \hline
        Message Reliability & Unreliable - the sender doesn't receive an ACK & Reliable - ACKs are sent on receipt of messages \\
      \hline
        Message Batching & Available out-of-box & Difficult to implement \\
      \hline
        Virtual Hosts & Not present & Used to ensure segregation of clients \\
      \hline
        Basic Distributed Unit & Topic & Queue \\
      \hline
        Message Delivery Model & Only pull model available & Both push and pull models are available \\
      \hline
        Consumer Subscription Model & Point-to-point and Pub-sub models available & Depending the type of exchange, both Point-to-point and Pub-sub models are implementable \\
      \hline
        Message Persistence & Writes to a persistent file system log using the page-cache & Durability is a configuration option while creating a queue \\
      \hline
        Application Domain & Log Aggregation, Big Data Analytics - high throughput, weaker durability & Financial services, Stock,  Banking - strong durability guarantees \\
      \hline
    \end{tabular}
    \caption{Message Broker Comparison} 
  \label{tab:message-broker-comparison}
  \end{table}


\section{Kafka}
\label{sec:kafka_appendix}

  \subsection{Kafka - Storage} 
  \label{sub:kafka_storage}
  \begin{itemize}
    \item 
    Each topic is a logical log. Physically, each log is a segment of a partition, each segment being of roughly the same size.
    \item 
    Flow of Data:
    \begin{itemize}
      \item 
      Producer publishes message
      \item 
      Broker appends message to the latest segment
      \item 
      Messages are buffered/batched before being flushed to the segment
      \item 
      Message is visible to the consumers after flushing
    \end{itemize}
    \item 
    No message IDs used. A message is identified as it's logical offset from the start of the logical log.
    \item 
    Kafka maintains in-memory, a list of the start of the segments that mark the start of each segment file.
    \item 
    Consumers are responsible for calculating the offset, and determine the next offset to request. Consumers can do a batched-pull if needed. Batched-pull would be done in terms of data-size, not number of messages.
    \item 
    No broker level caching is used. Kafka relies on OS-level page caching. `Write-through' and `Read-ahead' are the caching strategies that work well-enough. This is because, typically, a consumer only lags producers by a small amount of messages, hence, most messages are served from the broker.
    \item 
    The SendFile API is used to bypass the application and kernel buffers to write data directly from the file channel to the socket,
    \item 
    Brokers don't maintain state of any consumers. The data longevity on a broker is a time-based SLA configuration parameter. Consumers need to maintain their own current offset. This allows consumers to replay messages from a previous offset if needed.
  \end{itemize}
  

  \subsection{Kafka - Co-ordination} 
  \label{sub:kafka_co_ordination}

  \begin{itemize}
    \item 
    Consumer Groups are logically a single consumer, and can jointly consume messages published to a topic. Each message is read by only 1 consumer in the group.
    \item 
    The minimum unit of parallelism in Kafka is defined to be a partition.
    \item 
    There is always at least 1 consumer per partition of a topic. This implies that there is no co-ordination needed between consumers within a consumer group for a the last offset read from a partition.
    \item 
    Typically, $partitionCount > consumerCount $ 
    \item 
    Kafka doesn't use the concept of a leader/master node.
    \item 
    Zookeeper is used to provide a consensus service. The consensus to be arrived at is which consumers are the nodes that are participating as brokers and consumers, the topics that are currently active, and the offset of the last consumed message of each partition. Zookeeper is also used to trigger the re-balancing process for consumers when any change to the list of brokers and consumers are detected for a given topic.
    \item 
    Zookeeper uses a file-system like API, with data stored at paths. Paths can either be persistent or ephemeral (liveness depends on the creating client). It is fully replicated across the hosts chosen as Zookeeper nodes.
    \item 
    There are 4 types of registries store in Zookeeper, along with 
    \begin{itemize}
      \item 
      Broker: host name, port, set of topics and partitions stored
      \item 
      Consumer: Consumer group name, topics subscribed to.
      \item 
      Ownership: Mapping from a partition path to the path of the consumer that `owns' a particular partition.
      \item 
      Offset: Mapping from a partition path to the offset of the last message that was read from the partition.
    \end{itemize}
    \item 
    Each consumer registers a Zookeeper watcher on the brokers and consumers that correspond to the topics that it is currently subscribed to. If there is any change, then the re-balancing process is triggered.
  \end{itemize}

  \textbf{Re-balancing Process:}
  This is initiated when there is a change in the broker and consumer node list for a given topic.
  \begin{itemize}
    \item 
    Remove the currently subscribed partitions for each of the consumers in the consumer group from the Ownership registry.
    \item 
    List all the partitions for the topics that the consumer group is subscribed to.
    \item 
    Divide the partitions equally amongst the consumers, and update the Ownership registry.
    \item 
    Read the current offset for each of the assigned partition and resume reading and processing the messages from the point where it was left off.
    \item 
    Since there are situations in which the change notifications arrive at different times at different consumers in a group, a consumer might attempt to take ownership of a partition that another consumer has not relinquished ownership of yet. In this case, the first partition wait and retries.
  \end{itemize}        
  

  \subsection{Kafka - Delivery Guarantees} 
  \label{sub:kafka_delivery_guarantees}

  \begin{itemize}
    \item 
    Messages are relayed `at least once'.
    \item 
    2PC is required if `exactly once' semantics are needed, which is deemed to be overkill for Kafka.
    \item 
    If needed, the application needs to handle deduplication.
    \item 
    In-order delivery of messages is guaranteed from a single partition, due to the monotonically increasing offset of each log message.
    \item 
    CRCs are stored for each message to validate against data corruption. These CRCs are logged.
  \end{itemize}
  

\section{AMQP} 
\label{sec:amqp_appendix}

  \subsection{AMQP Components} 
  \label{sub:amqp_components}
    \begin{itemize}
      \item
      Broker: it is the server side of AMQP, receiving from producers and sending messages to right consumers.
      \item
      Producer: Client of AMQP who sends messages to a broker.
      \item
      Consumer: Client of AMQP who receives messages from brokers.
      \item
      Exchange: Part of the broker, receiving messages from producers, and routing them to  queues. It is created by clients.
      \item
      Queue: Part of the broker, receiving messages from exchanges and sending them to consumers. It is created by clients.
      \item
      Virtual host: Namespace for broker to specify the entities, such as exchanges and queues, application refers to. It is used to achieve access controls in broker. Clients need user account and password to get access to a virtual host in order to receive or send messages to broker. 
      \item
      Binding: Relationship between exchange and queue, created by clients with specified binding keys.
      \item
      Routing key: Key specified by a producer and included in messages.
    \end{itemize}
      

  \subsection{AMQP - Features}
  \label{sub:amqp-features}
    AMQP has some features that make it appealing to application developers. The most critical features include a choice of exchange types, message persistence, and distributed brokers.
    
    There are four types of exchanges defined in AMQP, including direct exchange, topic exchange, fanout exchange, and header exchange.
    \begin{itemize}
      \item 
      Direct exchange sends messages to different queues according to the routing key. Messages are load balanced between consumers instead of queues.
	  \item
      Topic exchange sends messages to all queues whose binding key matches routing key. It is used to implement publish/subscribe model.
      \item
      Fanout exchange acts like a broadcaster and does not consider routing keys. The exchange just sends received messages to all queues which are bound to the exchange.
      \item
      Header exchange uses message headers to route message to queues instead of routing key.
    \end{itemize}
    
    AMQP also offers strong reliability guarantees. For each exchange and queue in the broker, messages can be persisted in disk by toggling the durability flag. 
    
    In terms of distributed brokers, AMQP provides cluster, federation and shovel methods to achieve high availability and message brokers distribution. Nodes within the cluster are mirrors of each other, except queues which are local to each node. Federation and shovel architectures are used to allow messages across a WAN.

  \subsection{AMQP - Failure Handling} 
  \label{sub:amqp_failure_handling}
    When broker goes down and restarts, persistent queue, whose durable feature is set as true, will be recovered. However, only the persistent messages in the durable queue can be recovered. When the message is sent to the exchange, producer can assign delivery mode as 2 which means the message is persistent.
  


\section{Other Message Broker Implementation}
\label{sub:other_msg_broker_impl}
  \subsection{Apache ActiveMQ} 
  \label{sub:apache_activemq}
    \begin{itemize}
      \item 
      Built on top of the JMS specification, which an enterprise messaging standard included in Java EE.
      \item 
      Supports a variety of message broker protocols, including Stomp, AMQP and MQTT, in addition to traditional communications protocols like TCP, UDP and XMPP and also multiple topologies. The architecture is illustrated in Figure \ref{fig:activemq-architecture}.
      \begin{figure}[ht]
        \centering
        \includegraphics[width=.4\textwidth]{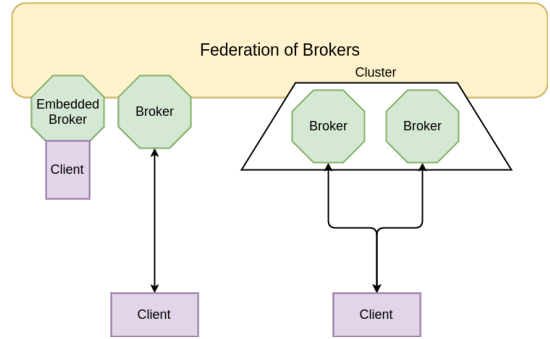}
        \caption{ActiveMQ Architecture}
        \label{fig:activemq-architecture}
      \end{figure}
      \item 
      ActiveMQ brokers can also be set up like a federated cluster, similar to RabbitMQ.
      \item 
      Active MQ is under active development after being incubated at Apache. There have been 79 commits to the source in the past 3 months. (beginning December 17, 2016). \footnote{https://git-wip-us.apache.org/repos/asf?p=activemq.git}
    \end{itemize}
  

  \subsection{ZeroMQ} 
  \label{sub:zeromq}  
    \begin{itemize}
      \item 
      Unlike traditional message queuing systems, ZeroMQ is broker-less.
      \item 
      "Smart endpoint, dumb network", was the philosophy behind ZeroMQ\cite{sustrik2015zeromq}
      \item 
      It was originally designed for stock-trading messages, with the objective of circumventing a central broker as a means of routing the messages.
      \item 
      Similar to ActiveMQ, it is a library, not a protocol. 
      \item 
      Figure \ref{fig:zeromq-archtecture} illustrates an example ZeroMQ architecture with node communicating between themselves without an arbiter, but merely a directory service to facilitate lookups.
      \begin{figure}[ht]
        \centering
        \includegraphics[width=.4\textwidth]{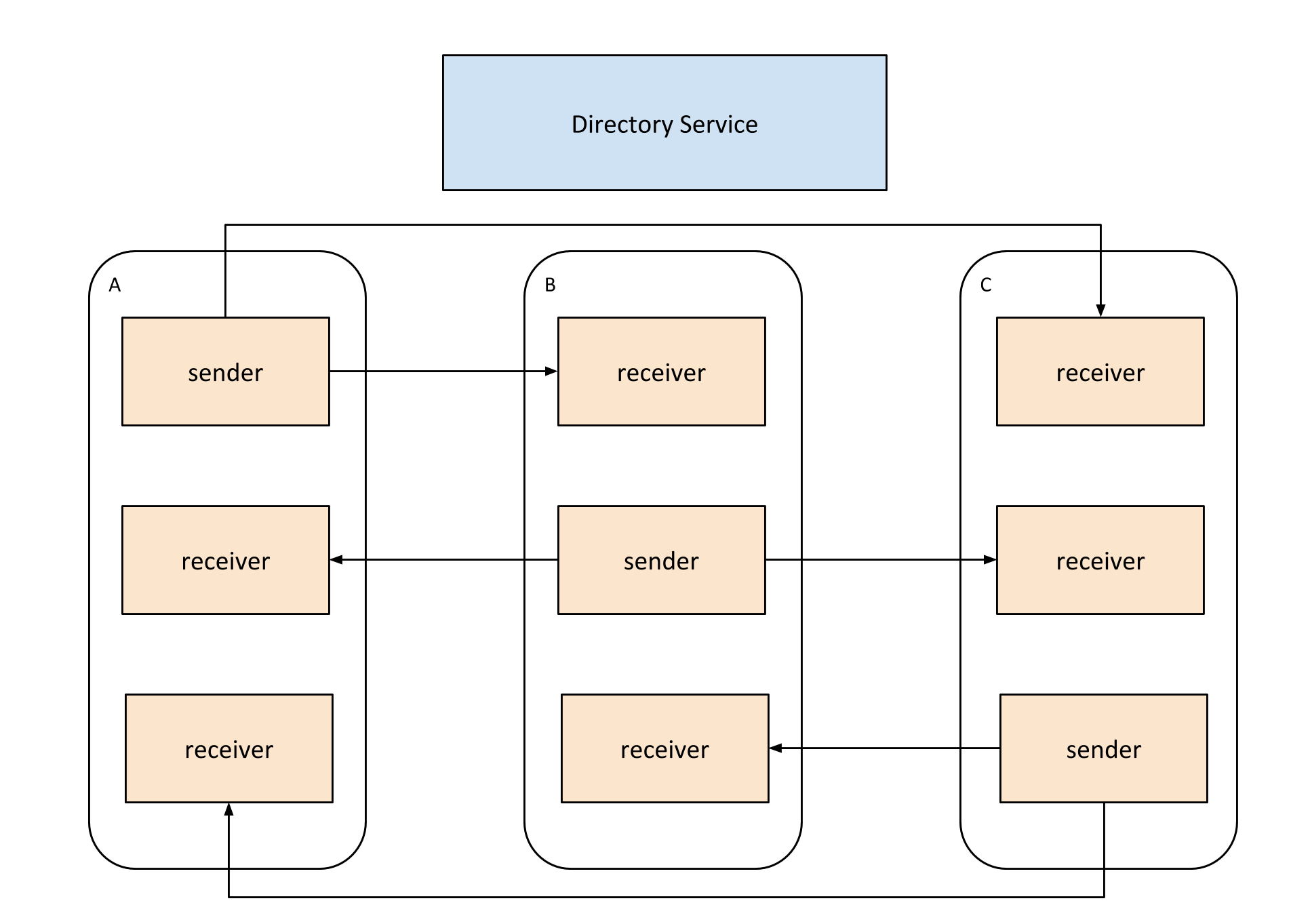}
        \caption{ZeroMQ Architecture}
        \label{fig:zeromq-archtecture}
      \end{figure}
      \item 
      Connections between ZeroMQ clients can be established with patterns like fan-out, pub-sub, task distribution, and request-reply.
      \item 
      ZeroMQ is under active development. There have been 143 commits to the source in the past 3 months. (beginning December 17, 2016). \footnote{git://github.com/zeromq/libzmq.git}
    \end{itemize}


  \subsection{RabbitMQ} 
  \label{sub:RabbitMQ}
    \begin{itemize}
      \item 
      RabbitMQ is a popular implementation of AMQP. It is developed by VM-Ware in Erlang.
      \item
      Compared to Apache Qpid, it is easy to use and deploy, and the code is short and clear.
      \item
      Some well-known RabbitMQ users include Mozilla, AT\&T and UIDAI.
      \item
      RabbitMQ community and supports are active. There have been 294 commits to the source in the past 3 months. (beginning December 17, 2016). footnote{https://github.com/rabbitmq/rabbitmq-server}
    \end{itemize}
  
  
  \subsection{Apache Qpid} 
  \label{sub:apache_Qpid}
    \begin{itemize}
      \item 
      It is an implementation of AMQP, and provides two  messaging broker version, one is in Java, the other is in C++.
      \item 
      It uses Corosync to achieve high availability. Similar to RabbitMQ, Qpid uses broker federation to achieve scaling of capacity. 
      \item
      It stores its queues in database or memory.
      \item 
      There have been 52 commits to the source in the past 3 months. (beginning December 17, 2016). \footnote{https://git-wip-us.apache.org/repos/asf/qpid-jms.git}
    \end{itemize}
  

  \subsection{Amazon SQS} 
  \label{sub:Amazon_SQS}  
    Amazon SQS is a message queuing service which provides scalable and highly-available queues to store messages so that they can be produced and consumed between components over the distributed system. It runs within Amazon high-availability data centers, so queues will be available whenever applications needs. All messages are stored redundantly across multiple servers and data centers, providing high availability. Amazon SQS provides standard queues in all regions and FIFO queue only in US West and East. Therefore, for most users, queue type they can use is standard queue, which delivers messages at least once and might deliver messages in different order from which messages were sent (best-effort ordering).

    The official website recommends four cases to use Amazon SQS\cite{AmazonWebServices2017}:
    \begin{itemize}
      \item 
      Tracking the success of each component of an application independently.
      \item
      Scheduling jobs in the queue with a delay individually.(up to 15 minutes with standard queues)
      \item
      Increasing concurrency or throughput dynamically at read time with no pre-provisioning.
      \item
      Scaling transparently as requests and load changes with no provisioning instructions.
    \end{itemize}
  
    
  \subsection{Microsoft Message Queuing (MSMQ)} 
  \label{sub:microsoft_message_queuing_}

    \begin{itemize}
      \item 
      Message Queuing (MSMQ)\footnote{https://msdn.microsoft.com/en-us/library/ms711472.aspx} technology enables applications running at different times to communicate across heterogeneous networks and systems that may be temporarily offline. 
      \item 
      Applications send messages to queues and read messages from queues. The following illustration shows how a queue can hold messages that are generated by multiple sending applications and read by multiple receiving applications.
      \item 
      Message Queuing provides guaranteed message delivery, efficient routing, security, and priority-based messaging.
      \item 
      It can be used to implement solutions to both asynchronous and synchronous scenarios.
    \end{itemize}

  
\end{document}